# Synergizing Knowledge Graphs with Large Language Models: A Comprehensive Review and Future Prospects

DaiFeng Li*
School of Information Management
Sun Yat-sen University
Guangzhou, China
lidaifeng@mail.sysu.edu.cn
*corresponding author

Fan Xu
School of Information Management
Sun Yat-sen University
Guangzhou, China
xufan9@mail2.sysu.edu.cn

*Abstract*—Recent advancements have witnessed the ascension of Large Language Models (LLMs), endowed with prodigious linguistic capabilities, albeit marred by shortcomings including factual inconsistencies and opacity. Conversely, Knowledge Graphs (KGs) harbor verifiable knowledge and symbolic reasoning prowess, thereby complementing LLMs' deficiencies. Against this backdrop, the synergy between KGs and LLMs emerges as a pivotal research direction. Our contribution in this paper is a comprehensive dissection of the latest developments in integrating KGs with LLMs. Through meticulous analysis of their confluence points and methodologies, we introduce a unifying framework designed to elucidate and stimulate further exploration among scholars engaged in cognate disciplines. This framework serves a dual purpose: it consolidates extant knowledge while simultaneously delineating novel avenues for real-world deployment, thereby amplifying the translational impact of academic research.

*Keywords—knowledge graphs, large language models, synergizing framework, comprehensive review*

## I. Introduction

In recent years, pre-trained language models (PLM) have witnessed rapid advancements, exemplified by influential architectures such as BERT[1], GPT[2], T5[3], GLM[4], all pre-trained on vast corpora. As computational resources have expanded, the scale of these models has escalated significantly, culminating in large language models (LLMs) like ChatGPT4 and LLaMA3, which exhibit remarkable proficiency across a spectrum of natural language processing (NLP) tasks[5-7]. Despite these strides, LLMs continue to face notable challenges:

1) Outdated Knowledge: Pre-trained LLMs struggle to incorporate up-to-date information, as their knowledge remains static and fails to capture the latest insights[8].

2) Parameter Rigidity: Fine-tuning these models necessitates high-quality data, entailing significant costs[9], and iterative fine-tuning can result in catastrophic forgetting.

3) Illusory Accuracy: LLMs occasionally produce responses that appear plausible but deviate from factual reality, manifesting as hallucinations[10, 11].

4) Lack of Transparency: Owing to their black-box nature, LLMs suffer from a dearth of interpretability[12], with knowledge implicitly embedded in model parameters and thus challenging to validate. The reasoning processes within deep neural networks remain opaque and difficult to elucidate.

Due to these constraints, significant challenges persist when deploying LLMs for specialized question answering tasks. Unlike LLMs, KGs encapsulate vast quantities of factual information in the form of structured triples, providing precise and explicit knowledge representations[13]. They are renowned for their symbolic reasoning capabilities[14], which yield interpretable outcomes. KGs are also capable of evolving as new knowledge is continually curated and incorporated by domain experts[15] to construct domain-specific graphs[16]. However, the construction of KGs remains a formidable task. Current methods for building KGs struggle with addressing their inherent incompleteness and managing the dynamic and complex nature of their evolution in practical applications. Additionally, these methods often overlook the rich semantic information embedded within KGs, limiting their versatility. Consequently, leveraging LLMs to address the challenges faced by KGs has become a pressing need. Recent efforts have coalesced around the integration of KGs into LLMs as a means to mitigate these limitations. This paper systematically reviews the pertinent research on the convergence of existing KGs and LLMs, and constructs a unifying framework by categorizing the methods of integration. Through our analysis, we aspire to catalyze innovative insights into the fusion of KGs and LLMs, and to advance the field of AI towards greater efficiency and trustworthiness.

## II. Relevant Concepts

### A. Essential Concepts for Knowledge Graphs

The notion of a Knowledge Graph (KG), introduced by Google in 2012, aims to enrich search engine capabilities. A Knowledge Graph is a structured knowledge base that models the real-world through a graphical representation of concepts, instances, relationships, rules, and events. This modeling is grounded in ontologies, which serve as shared conceptual schemas that define the terms and relationships within a domain.

Ontologies provide a formal description of concepts and their interrelations, acting as the foundational layer for most Knowledge Graphs[17]. They establish the conceptual and logical framework upon which Knowledge Graphs are built. While ontologies describe the abstract structure, Knowledge Graphs instantiate these structures with specific data. The factual content of a Knowledge Graph is typically encoded using triples, and the languages employed for this purpose are often advanced ontology languages, such as RDFS[18], OWL[19].

Knowledge mapping is a conceptual framework that links knowledge indices within a specific organization through a "map," thereby elucidating the types, characteristics, and interrelations of knowledge resources[20]. This mapping serves as a visual and sequential representation of knowledge lineage, used to depict both the evolution and the structural composition of knowledge[21].

The construction of a knowledge map primarily encompasses knowledge extraction, fusion, and processing. Knowledge extraction endeavors to automate or semi-automate the retrieval of entities, attributes, relations, and factual descriptions from unstructured, semi-structured, and structured data[22], leveraging information resource organization theories for management [23]. Advances in artificial intelligence and deep learning have significantly impacted joint knowledge relationship extraction, with models like LSTM-CRF, RNNs, Transformers, and BERT achieving notable success[24-29]. Event extraction, another key task, employs remote supervision and leverages pre-trained models for prompt and instruction learning [30]. A substantial volume of extracted knowledge requires integration, which typically involves knowledge disambiguation and alignment. State-of-the-art approaches often utilize sequence-based deep learning models to derive contextual embeddings for entities and apply graph theory along with unsupervised clustering techniques to achieve disambiguation and alignment[31, 32]. Knowledge processing also includes designing mechanisms for expert involvement, developing a conceptual system tailored to specific application scenarios, and structuring a knowledge backbone to refine the constructed knowledge graph[33-35].

### B. Conceptual Foundations of Large Language Models

Language models have evolved through four distinct phases [36]: statistical, neural, pre-trained, and large-scale. Statistical language models rely on the Markov assumption to predict word probabilities[37]. Neural language models, by contrast, employ neural networks to represent the likelihood of word sequences, incorporating the concept of word representations conditioned on aggregated context features[38]. Pre-trained models, exemplified by BERT, leverage the parallelizable Transformer architecture with self-attention[39]. They adopt a "pre-training and fine-tuning" paradigm, where the expansion of the model significantly boosts performance on downstream tasks[40]. As the number of parameters in pre-trained models grows, their performance across complex tasks improves according to a scaling law. For instance, GPT-3 demonstrates proficiency in solving few-shot tasks via contextual learning, a capability that GPT-2 lacks. This has led the academic community to refer to such large pre-trained language models as "Large Language Models (LLMs)".

LLMs refer to Transformer-based architectures with parameter counts in the hundreds of billions. These models are trained on vast corpora, such as those used for GPT-3[41], PaLM[42], LLaMA[43], GLM[4]. LLMs exhibit emergent capabilities[44], which are abilities that do not manifest in smaller models but arise in larger ones. Three notable emergent capabilities of LLMs include in-context learning, instruction following, and step-by-step reasoning. In-context learning (ICL), first formalized by GPT-3[41]. enables LLMs to generate expected answers by learning from word sequences in the input text without retraining or gradient updates. Instruction tuning on multi-task datasets allows LLMs to excel at previously unseen tasks, enabling them to follow new task instructions in a zero-shot setting, thus enhancing their generalization capabilities. Furthermore, LLMs can reason step-by-step, solving complex problems, such as mathematical questions, using a chain-of-thought approach[45].

### III. SURVEY METHODOLOGY

The systematic review conducted in this paper follows the methodology outlined by Kitchenham[46]. Our aim is to summarize and compare current approaches that harness LLMs to enhance the dynamism and versatility of KGs, mitigate the hallucination issues of LLMs through KG integration, and determine the strategies for leveraging KGs to improve the accuracy and interpretability of LLMs. To this end, we formulated a search query: "Knowledge Graph OR KG OR KGs AND Large Language Model OR LLM" for database searches across multiple platforms. The retrieved papers were deduplicated and thematically categorized to compile a comprehensive list of relevant studies, as illustrated in Figure 1. Additionally, we traced the references of the included papers to further expand our list of pertinent research literature.

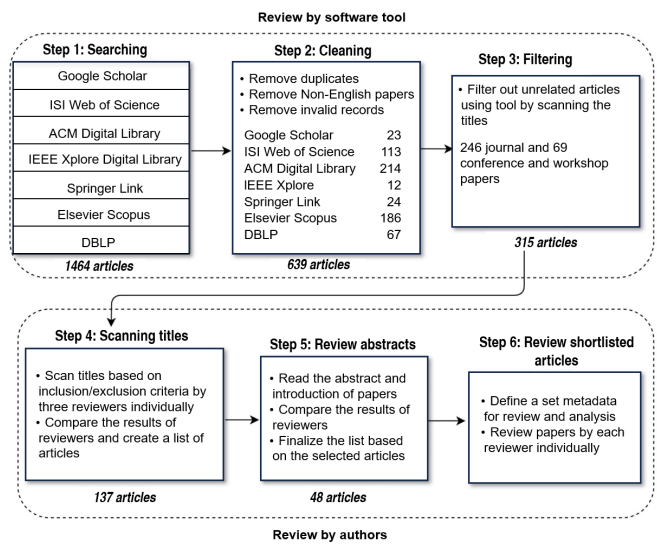

Fig. 1. The systematic review methodology

### IV. LITERATURE REVIEW

The synergy between KGs and LLMs fosters mutual enhancement. This interaction can be categorized into two primary areas: LLMs augmenting KGs and KGs enhancing LLMs.

*A. LLMs-augmented KGs*

According to the construction process of KGs, the enhancement of KGs by LLMs can be categorized into three primary areas: entity extraction, entity parsing and matching, and link prediction. We summarize and elaborate on the existing research in these domains.

*1) Entity Extraction:* Historically, the acquisition and deduction of relational knowledge have been anchored in symbolic knowledge bases, typically constructed through supervised extraction methodologies applied to unstructured textual corpora. Notably, the capacity of LLMs to retrieve, infer, and summarize relational facts in contexts such as question answering, via cloze-style prompts, or statement evaluation, has been lauded as a cardinal metric of their proficiency in comprehending and interpreting human language[47]. Despite the terminological diversity—encompassing knowledge or fact retrieval and reasoning—we herein designate the process of eliciting relational knowledge from LLMs as knowledge retrieval. As posited by Zhong [48] the model's fidelity might stem from rote memorization of training instances rather than genuine knowledge inference. Recent empirical evidence from KAMEL corroborates the significant gap between LLMs' abilities and accessing knowledge encapsulated within symbolic knowledge bases. Moreover, the Knowledge Memory, Identification, and Reasoning (KMIR) assessment underscores that the information retention capacity of LLMs is contingent upon the scale of their parameter space; smaller, compressed models exhibit enhanced memory preservation at the expense of diminished identification and reasoning prowess.

Cao[49] pioneered a tripartite taxonomy for eliciting factual knowledge from LLMs: prompt-based, case-based, and context-based methodologies. Among these, prompt engineering stands out as a cornerstone technique, meticulously crafting prompts to coax LLMs into producing responses tailored to specific tasks. LPAQA[50] exemplifying this approach, leverages automated mining and paraphrasing to forge a plethora of high-caliber prompts, subsequently aggregating answers derived from distinct prompts to enrich its knowledge base. Innovative strategies have also been proposed by scholars[47] harnessing LLMs like InstructGPT to generate initial instruction candidates. These are then refined through the incorporation of semantically akin instruction variants, a process that propels the system toward human-level performance in executing instructions. Early investigations into the grammatical nuances underpinning knowledge retrieval[51] revealed a pronounced reliance of both prompt structure and retrieval efficacy on grammatical integrity. Collectively, prior art has substantiated LLMs' capacity to apprehend relational knowledge to a notable degree. Yet, a critical gap persists regarding nuanced distinctions in performance across varied types of knowledge or relations—namely, the differential proficiency in handling common-sense knowledge compared to entity-centric encyclopedic facts, or horizontal relations juxtaposed against hierarchical ones. Moreover, concerns linger over LLMs' capability to extrapolate and infer knowledge beyond the confines of explicitly encountered statements, underscoring the need for further research in this domain.

*2) Entity Parsing and Matching:* Entity reconciliation and alignment constitute a pivotal endeavor in the realm of data science, entailing the meticulous linkage of disparate informational fragments scattered across a multitude of heterogeneous repositories, all converging upon a singular, underlying referential entity [52-54]. Historically, scholarly pursuits have fixated on the formulation of methodologies and the calibration of similarity metrics tailored to entities encapsulated within tabular, structured data formats. However, the advent of semi-structured data landscapes, epitomized by the burgeoning ecosystem of Knowledge Graphs, heralds a novel frontier in entity resolution. This paradigm shift necessitates an evolved approach, one that transcends the conventional confines of flat data structures. Entity alignment strategies, in response, have bifurcated into two primary paradigms: the traditional, generalist methods, and the more contemporary, embedding-centric techniques.

LLMs have emerged as versatile tools for the resolution and linking of entities within Knowledge Graphs (KGs), manifesting their utility through a myriad of applications[55]. Notably, LLMs offer a transformative approach to the laborious and time-demanding task of training data annotation—a bottleneck often encountered in the entity alignment of KGs. Analogous to the efficiency gains achieved through Generative Adversarial Networks (GANs) in reducing the manual effort required for data labeling [56], LLMs possess the potential to autonomously generate labeled KG samples, thereby outperforming traditional embedding-based methodologies in terms of efficacy and scale. Moreover, LLMs present an opportunity to streamline the construction of entity corpora, serving as a foundation for the development of matching rules—eschewing the traditional reliance on declarative formalizations encoded in logic languages within graph settings. To this end, LLMs can be fed with training data akin to logical language inputs, paralleling the usage of SQL statements on textual corpora. However, the expedited engineering of meaningful rule corpora for real-world, large-scale Knowledge Graphs, such as DBpedia[57] and Wikidata[58], remains a critical challenge. The application of entity matching rule logs to these expansive KGs mirrors the utility of query logs, underscoring the necessity of preparatory steps for comprehensive knowledge reasoning[59, 60]. In summation, the amalgamation of general entity linking techniques with embedding-based methods, augmented by the creation of rules and annotated data catalyzed by LLMs, paves the way for a more refined integration of LLMs with the intricate domain of knowledge reasoning[61]. This synergy not only enriches the capabilities of LLMs but also propels the advancement of KGs towards greater sophistication and utility in real-world applications.

*3) Link Prediction:* Link prediction involves forecasting the missing element of a triple given the other two components. This task includes predicting the head entity (?, r, t), the relation (h, ?, t), and the tail entity (h, r, ?). Most link prediction methods for knowledge graphs (KGs) focus on static snapshots of the graph. Many approaches involve a training phase where KGs are used to learn embeddings and other model parameters. However, these models often struggle to predict links for entities not seen during training. In contrast, Inductive Link Prediction (ILP) addresses the challenge of predicting links for new entities that were not part of the initial KG. Moreover, existing KG completion methods based on KG embeddings frequently underutilize textual and contextual information[62]. To address these limitations, recent research emphasizes integrating textual information from KGs to enhance embeddings and improve the performance of downstream tasks. Latent representations are derived from textual information using various encoding models, including linear models, convolutional models, recurrent neural models, and LLMs[63, 64]. KEPLER[65] offers a unified approach that integrates KG embeddings with pre-trained language representations, embedding both text-enriched and factual knowledge into LLMs. Nayyeri[66] utilizes LLMs to generate word, sentence, and document representations, which are then merged with graph structure embeddings. Huang[67] introduces a framework that integrates LLMs with other modalities to construct a unified multimodal embedding space.

## B. KGs-enhanced LLMs

According to the training and reasoning processes of LLMs and the integration points with KGs, related research primarily centers on the following areas: pre-training and fine-tuning of LLMs augmented with KGs; retrieval improvement of LLMs enhanced with KGs; and prompt refinement and reasoning of LLMs integrated with KGs.

*1) Pre-training and fine-tuning of LLMs integrated with KGs:* KGs typically contain information derived from highly trusted sources, post-processed, and vetted through human evaluation. Integrating KG information into pre-training corpora can mitigate the issue of limited information coverage inherent in the analyzed text[68, 69]. The primary approach for leveraging factual knowledge from KGs to LLMs involves explicitly injecting structured knowledge into the models[70]. The combination of KGs and LLMs, alongside efficient prompt design, facilitates the incorporation of structured knowledge and new, evolving information into LLMs[71], addressing the black-box nature of these models. Mainstream LLM fine-tuning techniques, such as Lora[72] and P-tuning[73] require high-quality, expert-annotated data in specific domains, which is often scarce. Consequently, these techniques face challenges in low-resource settings. The domain knowledge encapsulated in knowledge bases can serve as a high-resource knowledge representation to improve the effectiveness of low-resource language model tuning[74]. Studies have demonstrated that the foundational knowledge injected from KGs into LLMs yields significant performance enhancements in text generation and question-answering tasks[75]. An interpretable neural symbolic knowledge base was introduced[76], where the memory mechanism comprises vector representations of entities and relations drawn from the existing knowledge base. where the memory mechanism comprises vector representations of entities and relations drawn from the existing knowledge base.

*2) Retrieval enhancement of LLMs integrated with KGs:* Retrieval enhancement aims to improve the performance of information retrieval by leveraging LLMs, providing the most pertinent external knowledge to these models to generate more accurate answers[64]. Information retrieval technology is integral to search engines. Leading search engines can compute user query terms and generate a feature space at the billion level. They employ large-scale machine learning models, such as Logistic Regression (LR), Gradient Descent Boosting Trees (GDBT), and Deep Learning (DL), to retrieve and return the most relevant results through correlation calculations, learning-to-rank, and slot alignment[77]. Information retrieval optimization integrated with KGs remains a prominent research topic[78]. KGs can be utilized to generate knowledge features[79], facilitate deep semantic understanding[80], achieve knowledge alignment[81], and support explainable multi-hop reasoning[82]. For instance, "Who was the king of England during the American Revolutionary War?" is a query that requires multi-hop reasoning.

Integrating KGs with Information Retrieval systems provides the most pertinent information and structured knowledge from outside the LLMs, addressing the issue of knowledge cutoff. Knowledge cutoff refers to the LLMs' inability to perceive new knowledge and events that emerge after the training dataset. Incrementally loading new knowledge into LLMs through additional training is prohibitively expensive, as evidenced by the training costs of GPT-3 and PaLM. Research shows that during large-scale training, LLMs tend to favor popular, high-frequency common knowledge[83], while domain-specific expertise, including private and business-critical knowledge, is not well generated or applied in practical applications. Some relatively low-frequency, long-tail domain knowledge is not learned by LLMs[84]. One research direction to address the aforementioned knowledge gaps is knowledge editing: formulating optimization strategies through neuron detection and statistical analysis, and retraining and fine-tuning the model based on the modified data. However, retraining does not guarantee that incorrect data will be corrected[85]. Another strategy is to develop a hypernetwork to learn the parameter offset of the base model. For example, DeCao[86] trained a hypernetwork knowledge editor to modify the LLMs' erroneous understanding of a fact, using Kullback-Leibler divergence constraint optimization to mitigate side effects on other data/knowledge that should not be changed. However, this method performs poorly when multiple knowledge edits are made simultaneously, due to the use of the same strategy for multiple edits, which ignores the relationship between different edit gradients, leading to a "zero-sum" phenomenon where conflicts between gradients inevitably result in the failure of some data modifications. Based on these considerations, researchers have begun to introduce retrieval-generation architectures to construct retrieval-enhanced generative models. These methods primarily use unstructured paragraphs as external knowledge. Retrieval enhancement jointly trains the retriever and generator under the supervision of labeled answers[87]. For example, FiD concatenates the pre-trained external knowledge paragraph retrieval results and the original question, subsequently fusing them into the decoder for reasoning. However, due to the presence of interference noise in the paragraphs, this method exhibits reasoning biases in practical applications. Therefore, converting paragraph text into structured knowledge leads to better reasoning results. Structured knowledge becomes the primary source of external knowledge, and KGs can be used directly as external knowledge [88].

3) *Prompt optimization and reasoning of LLMs integrated with KGs:* Prompt engineering is a core technology for leveraging LLMs in specialized domains. It involves crafting and refining prompts to guide the output of AI models. Prompts serve as the interface between humans and AI systems, instructing the model on the type of response or task to perform[89]. The integration of KGs with LLMs has garnered increasing interest. Currently, research on combining KGs with prompt engineering has emerged as a prominent topic in the field. Some representative studies utilize KGs to automatically generate prompt templates. Compared to manually crafted templates, automatically generated prompts offer greater numbers, higher quality, and more diversity, while also considering meaningful learning patterns and having lower creation costs[90] with broader coverage. Other research explores how to integrate explicit knowledge from external sources, particularly through retrieval enhancement, to improve prompt engineering. This is achieved by providing additional context about entities through retrieval, enabling LLMs to generate more coherent reasoning paths[91]. Methods such as KnowPrompt[92] use KGs to incorporate semantic and prior knowledge between relationship labels into prompt tuning for relationship extraction, enhancing the prompt construction process and optimizing its representation through structured constraints. Representative studies like LARK have designed a prompt generation model with logical associations for reasoning tasks in KGs[93]. LARK employs retrieval techniques to identify entities and relations in queries, finding relevant subgraphs in KGs and forming associated contexts. It then utilizes LLM prompts that decompose logical queries to perform chained reasoning on these contexts. Experimental results demonstrate that this reasoning method significantly outperforms previous state-of-the-art models.

## V. Conclusion

In this paper, we present a framework for examining the methodologies and formalisms involved in the integration of KGs and LLMs. The framework was developed through a systematic review of recent literature, focusing primarily on the concepts and evolution of KGs and LLMs, LLM-augmented KGs, and KG-enhanced LLMs. Based on this review, we draw the following conclusions:

1) LLMs can assist in various stages of KG construction, encompassing entity discovery, coreference resolution, relation extraction, and even end-to-end KG construction and extraction from LLMs.

2) KGs can augment LLMs at different stages, including pre-training and fine-tuning, the inference phase, and prompt optimization, thereby mitigating the hallucination issue and improving the interpretability of LLMs.

3) The integration of KGs and LLMs has been applied in several domains, such as government, telecommunications, finance, and biomedicine, providing a theoretical foundation for practical application scenarios.

This work serves as a reference for future research endeavors by other scholars. Moreover, it highlights the potential for further investigation into the integration of KGs and LLMs in the following areas:

1) Hallucination Detection for LLMs Using KGs: Despite efforts to address the hallucination issue in LLMs, it is anticipated that this problem may persist. Consequently, leveraging KGs for hallucination detection in LLMs is likely to be a trending area of research.

2) Constructing Multi-Modal KGs with LLM Assistance: Current KGs primarily rely on textual and graph-based structures, whereas the real world encompasses multi-modal data. The advancement of LLMs in handling multi-modal data can facilitate the construction of multi-modal KGs.

Our findings can inform future research and development in the area of KGs and LLMs, contributing to the advancement of AI systems that better understand and represent complex real-world information.